\documentclass[aip,graphicx,amsmath,amssymb]{revtex4-1}
\usepackage{bm}
\draft 

\begin{document}


\title{Twisted $\mathbb{C}P^{N-1}$ instanton projectors and the $N$-level quantum density matrix} 



\author{Scott Shermer}
\email[]{scott.shermer@uconn.edu}
\affiliation{Department of Physics, University of Connecticut, Storrs, Connecticut 06269}


\date{\today}

\begin{abstract}
Twisted classical solutions to the $\mathbb{C}P^{N-1}$ model play a key role in the analysis of such models on the spatially compactified cylinder $\mathbb{S}_L^1 \times {\mathbb{R}^1}$ and have recently been shown to be important for the resurgent structure of this quantum field theory. Instantons and non-self-dual solutions both fractionalize, and domain walls formed by such topological solutions can be associated with $N$-vacua having maximally repulsive energy eigenvalues. The purpose of this paper is to reinforce this view through the investigation of a number of parallels between the $\mathbb{C}P^{N-1}$  model and $N$-level quantum mechanical density matrices. Specifically, we demonstrate the existence of a time-evolution equation for the $\mathbb{C}P^{N-1}$  instanton projector analogous to the Liouville-von Neumann equation in the quantum mechanical formalism. The group theoretical analysis of density matrices and the  $\mathbb{C}P^{N-1}$ model are also closely related. Finally, we explore the emergence of geometrical (Berry) phases in both systems and their interrelationship.
\end{abstract}

\pacs{}

\maketitle 


\section{Introduction}

It is well known that the two-dimensional $\mathbb{C}P^{N-1}$ model exhibits many features analogous to those possessed by four-dimensional non-Abelian gauge theories such as pure Yang-Mills.  One of the more salient features of this model is its possession of instanton solutions to first order self-dual equations characterized by integer toplogical charge \cite{Zak1, Zak2, Coleman}, as well as non-self-dual solutions with finite action \cite{Zak1, Zak2, Din} . In recent works, the physical significance of twisted boundary conditions has been explored for both instanton and non-self-dual solutions \cite{Eto1, Eto2, Bruckmann, Brendel, Dunne}.  More specfically, such fractionalized instantons arise in the context of the  $\mathbb{C}P^{N-1}$ mode with twisted boundary conditions and spatially compactified to the cylinder $\mathbb{S}_L^1 \times {\mathbb{R}^1}$ \cite{Bruckmann, Brendel, Dunne, Unsal, Dunne2}.  Domain walls formed by such topological solutions can be associated with $\theta$-vacua, and in fact the 1-loop effective potential for the twisted $\mathbb{C}P^{N-1}$ was derived in \cite{Unsal}.  This effective potential indicates the presence of $N$ vacua with maximally repulsive (evenly distributed) energy eigenvalues.  This view is bolstered by a number of recent results in toroidal $\mathbb{C}P^{N-1}$ models and topological insulators where topological properties can be ascertained from a Bloch wave analysis of the relevant systems  \cite{Thacker, Grusdt}.  The purpose of this paper is to reinforce this interpretation by expounding upon a number of previously uninvestigated parallels between the $\mathbb{C}P^{N-1}$ model and $N$-level quantum mechanical systems as represented by density matrices.  Both models possess a projector representation that can be defined in the complex projective space on a K\"ahler manifold.  In both models the parameters of the projective Hilbert space have Euler-angle type decompositions.  Specifically we draw parallels between the group theoretical analysis of QM density matrices \cite{Byrd1, Byrd2, Byrd3, Byrd4, Mukunda1, Mukunda2, Mukunda3} and a similar analysis of the $\mathbb{C}P^{N-1}$ model \cite{Mac1, Mac2, Mac3, Mac4}.  The time-evolution of the space of states in both models is determined by a Liouville-von Neumann equation which emphasizes the Hamiltonian-governed dynamics of each system as well as a connection between the first-order self-dual instanton equations and the Schr\"odinger equation.  Finally, both systems exhibit the emergence of a geometric (Berry) phase where, in the case of the field theory, the phase can be extrapolated from a decomposition of the Wilson loop operator \cite{Gubarev}.  These correspondences not only strengthen the interpretation of fractionalized instantons as tunneling events between classical vacua formed by the presence of a non-trival background field associated with twisted boundary conditions, but also serve to illuminate the potentially useful connections between the QM and field theoretical frameworks.

\section{Quantum Density Matrix Formalism: Review}

We begin with a review of the form and function of the density operator for an N-level quantum system \cite{Darius}.  For a QM system with a finite number of states $N$, we represent a statistical mixture of such states by the (positive Hermitian) density operator

\begin{equation}\rho  = \sum\limits_1^N {{p_n}\left| {{\psi _n}} \right\rangle \left\langle {{\psi _n}} \right|} \end{equation}
\\
where $\left| {{\psi _n}} \right\rangle$  is a complex one dimensional subspace (pure state) of the Hilbert space $\mathcal{H}$ spanned by all possible pure states, and the weights $p_n$ are subject to the constraint

\begin{equation}\sum\limits_1^N {{p_n} = 1}. \end{equation}
\\
Pure states are orthogonal projections represented by the density matrix satisfying the following conditions:

\begin{equation}\rho  = {\rho ^ \dag },{\text{    }}\rho {\text{ = }}{\rho ^2},{\text{   Tr}}\left[ \rho  \right]{\text{ = 1}}, \end{equation}
\\
whereas for a mixed state we allow $\rho \ne \rho^2$.  
\\
\indent In compact set notation we can denote the previous statements as follows:

\begin{align} \left\{ {{\text{pure}}} \right\}   \approx \left\{ {\rho  \in {\text{End}}\left( H \right)|{\text{ }}\rho  = {\rho ^ \dag },{\text{    }}\rho {\text{ = }}{\rho ^2},{\text{   Tr}}\left[ \rho  \right]{\text{ = 1}}} \right\} \\ \left\{ {{\text{mixed}}} \right\}  \approx \left\{ {\rho  \in {\text{End}}\left( H \right)|{\text{ }}\rho  = {\rho ^ \dag },{\text{    }}\rho  > 0,{\text{   Tr}}\left[ \rho  \right]{\text{ = 1}}} \right\}\end{align}
\\
where End$\left(\mathcal{H}\right)$ is the space of bounded linear operators $g$ mapping $\mathcal{H}$ to itself.  For the space of unit normalized pure states this is simply the projective Hilbert space $\bf{P}\mathcal{H}$ with 

\begin{align}\psi &=\lambda \psi  \\ \lambda \in \mathbb{C}&,\text{  } |\lambda|^2 =1 \\ \bf{P}\mathcal{H} := &\left(\mathcal{H} - \left\{0\right\}\right)/\mathbb{C}^* \end{align}
\\
The association of $\psi$ and $\lambda\psi$ is a consequence of the $U\left(1\right)$ gauge symmetry of the system, i.e., the phase of a pure state is not an observable property of the system.  
\\
\indent The density operator's time-evolution is governed by the Liouville-von Neumann equation

\begin{equation}i\frac{{\partial \rho }}{{\partial t}} =\left[ {H , \rho} \right]\end{equation}
\\
For a single level quantum system the corresponding projector is identified with the density matrix for the system.  For an N-level system, there is no unique representation for mixed states.  For a two-level system, the most familiar parameterization is the  Bloch sphere representation in which the density matrix is given in terms of the Pauli spin matrices by

\begin{equation}\rho  = \frac{1}{2}\left( {1 + \vec a \cdot \vec \sigma } \right).\end{equation}
\\
This parameterization implies the following:

\begin{align}
   \left| {\vec a} \right| & = 1{\text{  }} \Rightarrow {\text{  }}{\rho ^2} = \rho {\text{   Pure State}}  \\
   \left| {\vec a} \right| &< 1{\text{ }} \Rightarrow {\text{   }}{\rho ^2} \ne \rho {\text{   Mixed State}} \end{align}
\\
The Bloch sphere is thus a 2-sphere of radius $a$ in which points on the surface represent pure states, points interior to the surface represent mixed states, and antipodal points on the surface of the  represent mutually orthogonal state vectors.  Obviously the maximally mixed state lies at the center of the sphere.  The most common paramenterization for the two-state system is the Bloch parameterization in which the density matrix takes the following general form:

\begin{equation}\rho  = \left( \begin{matrix}
   a & 0  \\ 
   0 & {1 - a}  \\ 

 \end{matrix} \right).\end{equation}
\\
While this form of parameteriztion is generalizable to multi-level systems, it proves unwieldly as such, largely due to the non-rectangular nature of the integration domain for such generalizations \cite{Byrd4}.  Below we provide a more amenable trigonometric parameteriztion as presented by Byrd \cite{Byrd2, Byrd3}.
\\
\indent Before we present the parameterization, we proceed to make a few more observations about the general structure of a multilevel quantum system.  In the two-level case, the Hilbert space spanned by the state vectors is ${\mathbb{C}^2} \cong {\mathbb{R}^4}$.  Under the condition that the modulus squared be preserved, the unit vectors of the system will lie on the surface of the 3-sphere $S^3$ embedded in $\mathbb{R}^4$, and the physically measurable states are defined on this surface modulo a phase factor on $S^1$ as previously stated.  This reduces the space of physically realizable states to the 2-sphere as noted above, and the moding out of the phase is essentially the Hopf fibration

\begin{equation}{S^2} = {S^3}/{S^1}.\end{equation}
\\
What Byrd et al. observed is that normalization constraints and the association of states with different phases results in an isomorphism between pure state density matrices and $\mathbb{C}P^{N-1}$.  The Hilbert space of an $N$-Level system is $\mathbb{C}^N \approx \mathbb{R}^{2N}$.  This space is reduced to the $S^{2N-1}$ sphere under normalization constraints, and then to the complex projective space $\mathbb{C}P^{N-1}$ of the $2N-1$ sphere after the identification of states with different phases.\

\section{The $\mathbb{C}P^{N-1}$ Model}

For an introduction to and a more detailed treatment of the $\mathbb{C}P^{N-1}$ model, see the treatments by Zakrzewski and Coleman \cite{Zak1, Zak2, Coleman}.  The two-dimensional $\mathbb{C}P^{N-1}$ model is defined by $N$ spacetime fields and classical action density

\begin{equation} n = \left( {n_1}, \cdots,{n_N}\right)^T \end{equation}
\begin{equation} S=\int {{d^2}x{{\left( {{D_\mu }n} \right)}^\dag }\left( {{D_\mu }n} \right)} \end{equation}
\\
with a local $U(1)$ gauge symmetry $n \to {e^{i\alpha \left( {{x_1}} \right)}}n$.  The covariant derivative is defined in terms of a composite gauge field as follows:

\begin{equation}{D_\mu }  = {\partial _\mu }  -  i{A_\mu }\end{equation}
\begin{equation}{A_\mu } =-i {n^\dag }{\partial _\mu }n\end{equation}
\\
The fields are subject to the additional constraint

\begin{equation}n^\dag n =1\end{equation}
\\
The model consists of $N$ fields in $(1+1)$  spacetime dimensions $(x_0, x_1)$ and is, as it stands, highly nonlinear in the fields $n$ and $A_\mu$.  As previously mentioned, the model also exhibits a variety of non-perturbative phenomena the simplest of which are induced by a non-trivial topological charge characterized by $A_\mu$.  We can uncover this behavior by examining a particular subclass of the most general solutions.  This subclass satisfies the following first order identity along with the normalization condition (19)

\begin{equation}D_\mu n = \pm i\epsilon_{\mu\nu}D_\mu n \end{equation}
\\
The solutions to these first order equations are instanton solutions.  Equation (20)  assumes a particularly simple form if we introduce the projector representation

\begin{equation}{\mathbb{P}} \equiv n{n^\dag }\end{equation}
\\
satisfying the Hermitian projector conditions

\begin{equation}\mathbb{P}=\mathbb{P}^\dag=\mathbb{P}^2 \end{equation}
\\
After a transformation to holomorphic variables $z=x_0+ix_1$, (20) can be decoupled into either of the following pairs of equations (for instantons/anti-instantons, respectively):

\begin{equation}\mathbb{P}{\partial _ + }\mathbb{P} = 0;{\text{  }}{\partial _ + }\mathbb{PP} = 0\end{equation}
\begin{equation}\mathbb{P}{\partial _ - }\mathbb{P} = 0;{\text{  }}{\partial _ - }\mathbb{PP} = 0.\end{equation}
\\
The holomorphic derivatives are defined as

\begin{equation}
   {\partial _ + } \equiv \frac{\partial }{{\partial z}} = \frac{1}{2}\left( {\frac{\partial }{{\partial {x_0}}} - i\frac{\partial }{{\partial {x_1}}}} \right) \end{equation}
  \begin{equation} {\partial _ - } \equiv \frac{\partial }{{\partial \bar z}} = \frac{1}{2}\left( {\frac{\partial }{{\partial {x_0}}} + i\frac{\partial }{{\partial {x_1}}}} \right)  \end{equation}
\\
We now proceed to derive a commutator equation for the time evolution of the projector matrix in direct analogy with (9).  Since either side of the instanton equations above are vanishing, we see that the self-dual solutions in terms of projectors afford the exchange of spatial and time derivatives:

\begin{equation}\mathbb{P}{\partial _0}\mathbb{P} = -i\mathbb{P}{\partial _1}\mathbb{P}\end{equation}
\begin{equation}{\partial _0}\mathbb{P}\mathbb{P} = -i{\partial _1}\mathbb{P}\mathbb{P}\end{equation}
\\
With this observation we can recast the instanton equations in the following suggestive form:

\begin{equation}i \partial_0 \mathbb{P}\mathbb{P} =\left[ \mathbb{I}{\partial _1},{\mathbb{P}} \right]{\mathbb{P}}\end{equation}
\\
Upon comparison with the Liouville-von Neumann equation (9), we are led to associate the first term in the commutator with a Hamiltonian governing the time evolution of the $\mathbb{C}P^{N-1}$ projector via the Schr\"odinger equation.  For the ground state of the system, we assert that this projector evolution equation vanishes precisely for this reason.  The energy eigenvalues of the Hamiltonian are vanishing and suggest $N$ degenerate vacua for an $N \times N$ projector matrix.
\\
\indent We now proceed to introduce twisted boundary conditions on the compactified spatial coordinate as follows:

\begin{equation}n\left( {{x_0},{x_1} + \ell} \right) = {\Omega _0}n\left( {{x_0},{x_1}} \right)\end{equation}

\begin{equation} \Omega_0=\left( \begin{matrix}
e^{-2\pi i \mu _0} &  &  & 0 \\ 
 & e^{-2\pi i \mu _1} &  &   \\
  &  & \ddots &  \\
0 &  &  & e^{-2\pi i \mu_{N-1}} \\
\end{matrix} \right) \end{equation}

\begin{equation}\mu=\left(\mu_0,\mu_1,\cdots,\mu_{N-1}\right)=\left(0,1,2,\cdots,N-1\right)/N\end{equation}
\\
This is equivalent to a theory with periodic boundary conditions in the presences of a twisted $U(1)$ background field.  In terms of  periodic fields (for which an explicit representation is introduced in the next section), we can write

\begin{equation}{\tilde n_j}\left( {{x_0},{x_1}} \right) = {e^{ - i\frac{{2\pi {\mu _j}{x_1}}}{\ell}}}{n_j}\left( {{x_0},{x_1}} \right)\end{equation}

\begin{equation}{{\tilde n}_j}\left( {{x_0},{x_1} + \ell} \right) = {{\tilde n}_j}\left( {{x_0},{x_1}} \right)\end{equation}

\begin{equation} \Omega\left(x_1\right) = \left( \begin{matrix}
e^{\frac{-2\pi i \mu _0 x_1}{\ell}} &  &  & 0 \\ 
 & e^{\frac{-2\pi i \mu _1 x_1}{\ell}} &  &   \\
  &  & \ddots &  \\
0 &  &  & e^{\frac{-2\pi i \mu_{N-1} x_1}{\ell}} \\
\end{matrix} \right) \end{equation}
\\
\begin{equation}nn^\dag=\mathbb{P} \to \Omega \mathbb{P}{\Omega ^\dag }\end{equation}
\\
Thus, with the twisted boundary conditions the time-evolution equation for the projector matrix becomes

\begin{equation}i \partial_0 \mathbb{P}\mathbb{P} = \left[ \Omega^\dag{\partial _1}\Omega,{\mathbb{P}} \right]{\mathbb{P}}\end{equation}
\\
Evaluating the derivative in the commutator and following the line of reasoning presented for the untwisted case, we assert the following energy structure for the twisted system:

\begin{equation} H_{ij} \Leftrightarrow E_{ij}  \leftrightarrow  {\Omega ^\dag }{\partial _1}\Omega  =  \begin{pmatrix}
0 & & & &  & 0 \\
 &  1 & & & &  &\\
 & & 2 & & & & \\
& & & 3 & & &  \\
& & & &  \ddots & \\
0 & & & & & N-1 \\
\end{pmatrix} \end{equation}
\\
Throughout the rest of the paper, we will provide corroborating evidence to supplement these claims.

\section{Group Theory Relations Between the Density Matrix and $\mathbb{C}P^{N-1}$}

The space of possible states for the 2-level quantum system is represented by the coset spaces of $SU(2)$.  Cartan decomposition allows us to split the Lie algebra of $SU(2)$ into a Lie subalgebra $L(K)$ and the set of all other Lie algebra elements $U(P)$ \cite{Byrd2, Hermann}.  

\begin{equation}L\left( G \right) = L\left( K \right) \oplus U\left( P \right)\end{equation}
\\
The Lie algebra decomposition corresponds to the following decomposition of the Lie groups $K, P$:

\begin{equation}G = K \cdot P\end{equation}
\\
For $SU(2)$, we choose a one-parameter subgroup corresponding to $L(K)$ is 

\begin{equation}K = {e^{i a_2 {\sigma _2}}}\end{equation}
\\
A maximal subgroup of $SU(2)$ not containing $K$ is then

\begin{equation}S= {e^{i a_3 {\sigma _3}}}\end{equation}
\\
corresponding to the $U(1)$ subgroup.  We can now parameterize $P$ with the following construction

\begin{equation}P = KSK\end{equation}
\\
The full parameterization for $SU(2)$ becomes, explicitly, 

\begin{equation}U = {e^{i{\sigma _3}\alpha }}{e^{i{\sigma _2}\beta }}{e^{i{\sigma _3}\gamma }}\end{equation}
\\
where $\alpha, \beta, \gamma$ are the 3 euler-angle parameters with range $\alpha  \in \left[ {0,\pi } \right],\beta  \in \left[ {0,\pi /2} \right],\gamma  \in \left[ {0,2\pi } \right]$.  It is well known that this Euler angle parameterization leads to the following representation of the 2-level density matrix:

\begin{equation}\rho_{(N=2)}= \left( \begin{matrix}
\cos^2 \theta & 0 \\
0 & \sin^2 \theta \\
\end{matrix} \right)\end{equation}
\\
This can be demonstrated by examining the infinitesimal generator $K$ and applying a general rotation in $SU(2)$.

\begin{equation}x = \frac{1}{2}\left( {1 - { \sigma _2}} \right)\end{equation}
\begin{equation}x' = Ux{U^{ - 1}}\end{equation}
\\
\indent This is how we obtain the represetation for an arbitrary Bloch vector as in (10).  Such a rotation essentially defines the Hopf fibration to the complex projective space of states.  

\begin{equation}U{\sigma _i}{U^{ - 1}} = {R_{ij}}{\sigma _j}\end{equation}
\begin{equation} \Rightarrow \end{equation}

\begin{equation}x = \frac{1}{2}\left( {1 - {R_{2j}}{\sigma _2}} \right) = \frac{1}{2}\left( {1 + {a_j}{\sigma _j}} \right)\end{equation}
\\
Writing $n= \psi \sigma_i \psi^\dag$, it can be shown that

\begin{equation} \psi = e^{i\chi} \left( \begin{matrix}
e^{i \phi} \sin{\theta}\\
\cos{\theta} \\
\end{matrix} \right)\end{equation}
\\
The pure state density matrix (45) is just the diagonal of $\psi \psi^\dag$, and the $\chi$ phase is the unobservable $U(1)$ gauge as previously discussed.  We can obtain a general mixed density matrix by applying a unitary transformation in the complex projective space

\begin{equation}\rho  \to \rho  = U\rho {U^{ - 1}} = U\rho {U^\dag }\end{equation}
\\
The same procedure can be applied to the 3-level density matrix which takes the form

\begin{equation}\rho_{(N=3)}= \left( \begin{matrix}
\cos^2 \theta_1 \sin^2 \theta_2 & 0 & 0 \\
0 & \sin^2 \theta_1 \sin^2 \theta_2 & 0 \\
0 & 0 & \cos^2 \theta_2 \\
\end{matrix} \right)\end{equation}
\\
Cartan decomposition for the $SU(3)$ Lie group yields

\begin{equation}L\left( G \right) = L\left( K \right) \oplus U\left( P \right)\end{equation}
\\
In infinitesimal form, the generator belonging to the $SU(2) \otimes U(1)$ subgroup $K$ is

\begin{equation}U = 1 + \frac{i}{2}\left( {\sum\limits_{a = 1}^3 {{\varepsilon _a}{\lambda _a} + \eta {\lambda _8}} } \right)\end{equation}
\\
where the $\lambda_i$ are the Gell-Mann matrices, a 3-dimensional generalization of the Pauli matrices (such generalizations exist for any dimensionality).  The generator of transformations not belonging to this subgroup (i.e. the generator of transformations that belong to $P$) is

\begin{equation}U = 1 + \frac{i}{2}\left( {\sum\limits_{a = 4}^7 {{\varepsilon _a}{\lambda _a}} } \right)\end{equation}
\\
We choose a one-parameter subgroup of $P$

\begin{equation}S = {e^{i{\lambda _5}\theta }}\end{equation}
\\
Since $\lambda_8$ commutes with the $SU(2)$ subgroup $K$, we can write an element of this subgroup as 

\begin{equation}K = {e^{i{\lambda _3}\alpha }}{e^{i{\lambda _2}\beta }}{e^{i{\lambda _3}\gamma }}{e^{i{\lambda _8}\phi }}\end{equation}
\\
Using (43), a generic element of $SU(3)$ can then now expressed as

\begin{equation}U = {e^{i{\lambda _3}\alpha }}{e^{i{\lambda _2}\beta }}{e^{i{\lambda _3}\gamma }}{e^{i{\lambda _5}\theta }}{e^{i{\lambda _3}a}}{e^{i{\lambda _2}b}}{e^{i{\lambda _3}c}}{e^{i{\lambda _8}\phi /\sqrt 3 }}\end{equation}
\\
with Euler angle ranges 

\begin{equation}\alpha ,\gamma ,a ,c \in \left[ {0,\pi } \right)\end{equation}
\begin{equation}\beta ,\theta ,b \in \left[ {0,\pi /2} \right]\end{equation}
\begin{equation}{\phi} \in \left[ 0,\sqrt 3  \pi \right)\end{equation}
\\
Following the same logic as for the 2-level system, we can start from an infinitesimal generator and rotate to find the general vector representation for a 3-level system.

\begin{equation}x' = U\frac{1}{3}\left( {1 - \sqrt 3 {\lambda _8}} \right) U^{-1}\end{equation}
\\
We obtain

\begin{equation}U{\lambda _i}{U^{ - 1}} = {R_{ij}}{\lambda _j}\end{equation}
\begin{equation}x = \frac{1}{3}\left( {1 - \sqrt 3 {R_{8j}}{\lambda _j}} \right) = \frac{1}{2}\left( {1 + {n_j}{\lambda _j}} \right)\end{equation}
\\
The components of $\vec n$ are given by Byrd \cite{Byrd3}, yielding

\begin{equation}n= \psi \lambda_i \psi^\dag\end{equation}

\begin{equation} \psi = e^{i\chi} \left( \begin{matrix}
e^{i \phi_1} \sin{\theta_1} \cos {\theta_2} \\
e^{i \phi_2} \sin{\theta_1} \sin {\theta_2} \\
\cos{\theta_1} \\
\end{matrix} \right)\end{equation}
\\
Again, the pure state density matrix (53) is obtained by writing down the diagonal elements of $\psi \psi^\dag$ as can be verified.
\\
\indent We now turn to the "Euler angle" parameterization of the $\mathbb{C}P^{N-1}$ model.  A complex vector of any dimension can be parameterized by a set of complex hyperspherical coordinates $(\vec \theta , \vec \phi)$ in the following manner:

\begin{equation} \begin{pmatrix} 
  {n_1} \\
  {n_2} \\
  {n_3} \\ 
   \vdots  \\
  {n_N} \\
 \end{pmatrix} = \left( \begin{array}{l}
  {e^{i{\varphi _1}}}\cos \theta_1 \hfill \\
  {e^{i{\varphi _2}}}\sin \theta_1 \cos  \theta_2 \hfill \\
  {e^{i{\varphi _3}}}\sin \theta_1 \sin  \theta_2 \cos \theta_3 \hfill \\
   \vdots\hfill \\
  {e^{i{\varphi _N}}}\sin  \theta_1 \sin  \theta_2 \sin  \theta_3 \cdots \sin  \theta_{N-1}  \hfill \\
 \end{array} \right)  \end{equation}

\begin{equation}{\theta _i} \in \left[ {0,\pi /2 } \right],{\varphi _i} \in \left[ {0,2\pi } \right)\end{equation}
\\
Following Dunne, Unsal \cite{Unsal}, we elect to represent the fields of the $\mathbb{C}P^{N-1}$ model this way, we see that the diagonal elements of the projector $\mathbb{P}=nn^\dag$ become

\begin{equation}\mathbb{P}_{(N=2)}= \left( \begin{matrix}
 \cos^2 \theta & 0  \\
 0 & \sin^2 \theta \\
\end{matrix} \right)\end{equation}
\\
\begin{equation}\mathbb{P}_{(N=3)}= \left( \begin{matrix}
 \cos^2 \theta_2 & 0 & 0 \\
0 & \cos^2 \theta_1 \sin^2 \theta_2  & 0 \\
0 & 0 & \sin^2 \theta_1 \sin^2 \theta_2 & 0 \\
\end{matrix} \right)\end{equation}
\\
in direct analogy with (45), (53) after the swapping of angles ${\theta _1} \leftrightarrow {\theta _2}$ and a permutation of matrix elements for the $N=3$ case.
\\
\indent With respect to Cartan decomposition, generalized $\mathbb{C}P^{N-1}$ groups are characterized by the coset space 

\begin{equation}\mathbb{C}{P^{N - 1}} = \frac{{U\left( N \right)}}{{U\left( 1 \right) \times U\left( {N - 1} \right)}}\end{equation}
\\
which is just a special instance of Grassmanian manifolds characterized by the coset

\begin{equation}G\left( {n, r} \right) = \frac{{U\left( n \right)}}{{U\left( r \right) \times U\left( {n - r} \right)}}\end{equation}
\\
Hence,  $\mathbb{C}P^{N-1}$ is manifested by a Grassmanian with $n=N-1$ and $r=1$ \cite{Byrd1}.  The permissible parameterizations of such cosets are explored in detail by Gilmore \cite{Gilmore}.  Elsewhere, Macfarlane applies such techniques to an extension of  the $\mathbb{C}P^{N-1}$ projector representation to any $N$ \cite{Mac1, Mac2}.  Gilmore gives the general projective coordinates of such an $n \times r$ coset space as

\begin{equation} \text{Exp} \left(  \begin{matrix}
 & & & \vline & b_1 \\
& & & \vline & b_2 \\
& & & \vline &  \vdots \\
& & & \vline& b_n \\ \hline
b^\dag & & & \vline & \\
\end{matrix} \right)
\to \left(  \begin{matrix}
 & &  & \vline  & x_1 \\
& & & \vline & x_2 \\
& & &  \vline &  \vdots \\ 
& & &  \vline & x_n \\ \hline
x^\dag & & & \vline & r_{n+1}=(1-x^\dag x)^{1/2}  \\
 \end{matrix} \right) \end{equation}
\\
If we make the following decomposition of a general $n \times r$ column vector (with $x=KH$):

\begin{align} n & = \left( \begin{matrix}
1 & K \\
-K^\dag & 1\\
\end{matrix}\right) \left( \begin{matrix}
L & 0 \\
0 & H \\
\end{matrix} \right) \\
& = \left( \begin{matrix}
L & KH \\
-K^\dag L & H \\
\end{matrix} \right) \end{align}
\\
subject to the constraints (to enforce $nn^\dag =1$)

\begin{align} H^2 + KL^2K  = 1 \\
HK  = KL \\
\left(K^\dag K +1\right)L^2 & =1 
\end{align}
\\
we can generalize the $\mathbb{C}P^{N-1}$ system to other algebras through the use of quaternionic parameters \cite{Mac1, Mac2, Gilmore}.
\\
For example, take the element

\begin{equation}n_1 = n \left( \begin{matrix}
0 \\
1 \\
\end{matrix}\right)\end{equation}
\\
We then obtain the following projector representations:

\begin{equation}\mathbb{P}_1=n_1 n^\dag_1 = \left( \begin{matrix}
KH^2K^\dag & KH^2 \\
H^2 K^\dag & H^2 \\
\end{matrix} \right) \end{equation}

\begin{equation}\mathbb{P}^2_1 = \mathbb{P}_1\end{equation}

\begin{equation}\mathbb{P}_2=n_2 n^\dag_2 = \left( \begin{matrix}
L^2 & -L^2K \\
- K^\dag L^2 & K^\dag L^2 K \\
\end{matrix} \right) \end{equation}

\begin{equation}\mathbb{P}^2_2 = \mathbb{P}_2\end{equation}
\\
where in this case the two Hermitian projectors are also orthogonal

\begin{equation}\mathbb{P}_1\mathbb{P}_2 = 0\end{equation}

\section{Wilson Loops and Effective Potentials in the $\mathbb{C}P^{N-1}$ Model}

We now turn to an analysis of the role of the gauge potential in determining the dynamics of $\mathbb{C}P^{N-1}$ model.  In general, the relationship

\begin{equation}    i \frac{\partial}{\partial t} \psi  = A\psi  \end{equation}
\\
immediately allows us to write ($\rho=\psi \psi^\dag$)
 \begin{equation} i\frac{\partial }{\partial t}\left[ {\psi {\psi ^\dag }} \right] = \left[ {A,\psi {\psi ^\dag }} \right]  \end{equation}
\\
and thereby identify the gauge potential with the Hamiltonian.
\\
\indent For the case of thermal compactification, the path-ordered Wilson loop $W(x_0)$ is defined as a solution of the first order differential equation:

\begin{equation}\left( {i{\partial _0} + A} \right)\left| \psi  \right\rangle  = 0\end{equation}

\begin{equation}\left| {\psi \left( {{x_0}} \right)} \right\rangle  = W\left( x_0 \right)\left| {\psi \left( 0 \right)} \right\rangle \end{equation}
\\
where $W(x_0)$ is explicitly expressed as

\begin{equation}W\left( {{x_1}} \right) = P\exp \left\{ {i\int_0^T {A\left( {{x_0},{x_1}} \right)d{x_0}} } \right\}\end{equation}

\begin{equation}A\left( {{x_0}} \right) = A_\mu ^a\left( {{x_0},{x_1}} \right){{\dot x}_\mu }{T^\alpha }\end{equation}
\\
Here the $T^\alpha$ denote the gauge group generators for a given representation \cite{Gubarev}.  As Gubarev et al. observed, in this case we can associate the Wilson loop as the Schr\"odinger time-evolution operator for a background time-dependent potential $-A(x_0)$.  Recall however that the instanton equations of the $\mathbb{C}P^{N-1}$ model allow us to exchange the role of time and space derivatives so that in the spatially compactified model we obtain the following first order differential equation and Wilson loop operator:

\begin{equation}\left( { - {\partial _1} + A} \right)\left| \psi  \right\rangle  = 0\end{equation}

\begin{equation}\left| {\psi \left( {{x_1}} \right)} \right\rangle  = W\left( {{x_1}} \right)\left| {\psi \left( 0 \right)} \right\rangle \end{equation}

\begin{equation}W\left( {{x_0}} \right) = P\exp \left\{ {i\int_0^L {A\left( {{x_0},{x_1}} \right)d{x_1}} } \right\}\end{equation}
\\
Thus we see that the spatially compactified model can be viewed as being governed by a spatially varying background gauge potential $+A(x_1)$.
\\
\indent Dunne and Unsal have shown \cite{Unsal} that the $\mathbb{C}P^{N-1}$ 1-loop effective potentials for both the anti-periodic thermally compactified $(-)$ and the twisted spatially compactified models $(+)$, respectively, take the form

\begin{equation}{V_ - }\left| {W\left( {{x_1}} \right)} \right| = \frac{2}{{\pi {\beta ^2}}}\sum\limits_{n = 1}^\infty  {\frac{1}{{{n^2}}}} \left( {1 - {\text{Tr}}\left| {W\left( {{x_1}} \right)} \right|} \right)\end{equation}

\begin{equation}{V_ + }\left| {W\left( {{x_0}} \right)} \right| = \frac{2}{{\pi {L^2}}}\sum\limits_{n = 1}^\infty  {\frac{1}{{{n^2}}}} \left( {1 - {\text{Tr}}\left| {W\left( {{x_0}} \right)} \right|} \right).\end{equation}
\\
They also showed \cite{Unsal} that minimizing these effective potentials results in the following degenerate energy spectrum for the thermal case:

\begin{equation} V_{- \text{min}}= e^{i\frac{2\pi k}{N}} \left( \begin{matrix}
1 &  &  & 0 \\ 
 & 1 &  &   \\
  &  & \ddots &  \\
0 &  &  & 1 \\
\end{matrix} \right) \end{equation}
\\
and the following maximally repulsive spectrum for the spatially compactified and twisted model:

\begin{equation} V_{+ \text{min}}=\left( \begin{matrix}
e^{-2\pi i \mu _0} &  &  & 0 \\ 
 & e^{-2\pi i \mu _1} &  &   \\
  &  & \ddots &  \\
0 &  &  & e^{-2\pi i \mu_{N-1}} \\
\end{matrix} \right) \end{equation}
\\
Thus a connection is bridged between 
\begin{align*} &\text{1)  the twisted  $\mathbb{C}P^{N-1}$ projector evolution equation (37)} \\
& \text{2) the Wilson loop for the twisted }  \mathbb{C}P^{N-1} \text{model}\\
& \text{3) the energy spectrum implied by each, respectively.}\end{align*}
\\
This connection serves as further confirmation of our original interpretation of the projector matrix time-evolution equation as the $\mathbb{C}P^{N-1}$ model equivalent to the Liouville-von Neumann equation for the density matrix in ordinary quantum mechanics.  In the QM case, the evolution of the density matrix is determined explicitly in terms of the Hamiltonian dynamics of the system.  For a time-independent configuration, the Schr\"odinger equation gives the obvious link between the multi-level Hamiltonian and the energy eigenvalues of that system.  For the twisted $\mathbb{C}P^{N-1}$ model, we interpret equation (37) as an invitation to analyze the energy minima of the model as a spectrum of eigenvalues induced by the background potential associated with the twist.  Furthermore, we assert that the fluctuations around and tunneling between these minima are embodied by the fractionalized instantons investigated in Dabrowski et al. and Misumi, et al. \cite{Dunne, Misumi}.  We also note here that a similar interpretation has recently been arrived at for non-compactified models \cite{Thacker}.  Here the background gauge induces a discrete set of $\theta$-vacua and tunneling between these vacua is effectively formulated in terms of Bloch wave eigenstates of the gauged Hamiltonian.

\section{Berry Phase and Wilson Loops in $N$-Level QM and $\mathbb{C}P^{N-1}$}

The discussion of the previous section naturally leads us to our final consideration.  Multi-level QM systems can exhibit the emergence of both Abelian and non-abelian Berry phases associated solely with the geometry of the projective space of states of the system \cite{Byrd1, Byrd2, Byrd3, Byrd4, Mukunda1, Mukunda2, Duzzioni}.  In light of the previously emphasized similarities between the $N$-level QM system and the $\mathbb{C}P^{N-1}$ model, then, we should also expect that the field theory exhibits some sort of geometric phase.  Following the observations of Thacker and Gubarev, et al.\cite{Thacker, Gubarev}, we will proceed to show that this is indeed the case--  the Wilson loop provides the natural generalization of Berry phase from the QM case to field theory.
\\
\indent Both Abelian and non-Abelian geometric phases can arise in QM systems in the context of the density matrix formalism.  As a reminder, the Berry phase is a phase induced by the evolution of a quantum system through the space of possibly states and distinct from the dynamical phase induced directly by the Hamiltonian \cite{Berry, Simon, Shapere}. The Berry phase is purely determined by the geometry of the space of states irrespective of the dynamics of the system.  Berry's original deriviation appealed to adiabatic evolution of a state around a closed path, but the concept of geometric phase has since been generalized and the basic assumptions of adiabaticity and closed path circulation have become unnecessary. Expressions for the Berry phase for the 2-level and 3-level QM systems have been derived by Byrd, et al. \cite{Byrd1, Byrd2, Byrd3}.  We have already seen that these systems can be viewed as living in the space of $\mathbb{C}P^{1}$  and $\mathbb{C}P^{2}$, respectively.  The general expression for the Berry phase in these systems is

\begin{equation}{\varphi _g} = i\int \mathcal{A} \end{equation}

\begin{equation}\mathcal{A} = {\mathcal{A}_\mu }d{x^\mu } =   - i\psi d\psi  \end{equation}
\\
Using the expressions for $\psi$ given in (51) and (67), we can explicitly calculate the geometric phase induced along a given path.  The results are given by Byrd\cite{Byrd3}.  Our point here is to draw a parallel with the discussion surrounding the Wilson loop operator in the previous section.  Byrd also shows that the following unitary matrix determines the evolution of a 3-level system with a two-fold degeneracy\cite{Byrd1}:

\begin{equation}\psi \left( T \right) = U\left( n \right)\psi \end{equation}

\begin{equation}U\left( n \right) = {e^{\frac{i}{\hbar }\int_0^T {{E_n}\left( t \right)} }}P\left[ {{e^{i\oint_C \mathcal{A} }}} \right]\end{equation}
\\
where $T$ denotes the period of the system, $E_n$ its energy eigenvalues, and $C$ is a closed path on the parameterized manifold.  The first exponential corresponds to the dynamical phase and is determined by the details of the Hamiltonian.  The second term is the geometric phase and should be compared with (88), (92).  This comparison is precisely that made By Gubarev, et al. \cite{Gubarev}.  In fact, a generalized method for computing Abelian and non-Abelian geometric phases for open paths non-adiabatic paths was given by Duzzioni, et al. \cite{Duzzioni}.  A scheme for measur g these phases was recently proposed in the context of topological insulators and optical lattices \cite{Grusdt}.  The authors therein identify the use the Wilson loop approach to predict a non-Abelian Berry phase as measurable via a phase shift in the Bloch wave functions across the Brillouin zone.  This serves as a more down to earth corroboration with Thacker's interpretation of Bloch wave tunneling between $\theta$-vacua and our interpretation of the twisted $\mathbb{C}P^{N-1}$ model as being characterized by minima with maximally repulsive eigenvalues between which fractionalized instanton tunneling events occur.  Furthermore, the explicit expression for the connection one form in terms of the Wilson loop for the $\mathbb{C}P^{N-1}$ model looks locally like the results of Byrd et al. in terms of the complex hyperspherical parameterization.

\section{Conclusion}

In this paper we have elucidated a number of structural correspondences between the twisted $\mathbb{C}P^{N-1}$ field theory and N-level quantum mechanical density matrices.  Such comparisons yield novel perspectives and techniques for analyzing the vacuum structure of the two dimensional field theory.  The hyperspherical parameterization proves to be a useful tool in both the density matrix context and in the $\mathbb{C}P^{N-1}$ field theory.  Exploration of the mathematical structure of density matrices has and continues to be a fruitful and active area of research, most recently in the context of quantum computation.  We believe that salient features inherent to such quantum mechanical systems may have immediate and heretofore overlooked consequences, and we show here that it also provides a natural framework for field theoretical analysis.  The Berry phase is one of the most fundamental features common to both systems, and the geometrical nature of the phenomenon implies a type of universality with implications for the vacuum structure of field theories.  In the context of $\mathbb{C}P^{N-1}$,  the most imminent questions raised by this paper are related to the structure of the Liouville von-Neumann type projector evolution equation as well as the detailed form of the field theoretical Wilson loop/Berry phase.


\begin{acknowledgments}
The author would like to thank the OMSP fellowship program at University of Connecticut for supporting this work.  The author also thanks Gerald Dunne for comments and suggestions related to this work.
\end{acknowledgments}


\end{document}